\documentstyle[spie]{article}  
\input{psfig}    
\input{epsf}

\def\deg{\nobreak{$^\circ$}}
\def\arcmin{\nobreak{$'$}}
\def\arcsec{\nobreak{$''$}}
\newcommand{\um}{\,\hbox{$\mu$m}}
\newcommand{\mv}{\hbox{m$_{\rm V}$}}
\newcommand{\mk}{\hbox{m$_{\rm K}$}}
\newcommand{\av}{\hbox{A$_{\rm V}$}}
\newcommand{\msun}{\,\hbox{M$_{\odot}$}}
\newcommand{\lsun}{\,\hbox{L$_{\odot}$}}
\newcommand{\brg}{Br$\gamma$}
\newcommand{\hiiregion}{\hbox{H\,{\sc ii}~region}}
\newcommand{\pcmsq}{\,\hbox{cm}$^{-2}$}
\newcommand{\pcmcu}{\,\hbox{cm}$^{-3}$}
\newcommand{\kms}{\,\hbox{\hbox{km}\,\hbox{s}$^{-1}$}}
\def\h2{H$_2$}
\def\s1{1-0\,S(1)}
\def\spose#1{\hbox to 0pt{#1\hss}}
\def\simlt{\mathrel{\spose{\lower 3pt\hbox{$\mathchar"218$}}
     \raise 2.0pt\hbox{$\mathchar"13C$}}}
\def\simgt{\mathrel{\spose{\lower 3pt\hbox{$\mathchar"218$}}
     \raise 2.0pt\hbox{$\mathchar"13E$}}}
\def\eps@scaling{.95}
\def\epsscale#1{\gdef\eps@scaling{#1}}
\def\plotone#1{\centering \leavevmode
    \epsfxsize=\eps@scaling\columnwidth \epsfbox{#1}}
\def\plotfiddle#1#2#3#4#5#6#7{\centering \leavevmode
    \vbox to#2{\rule{0pt}{#2}}
    \includegraphics{#1}}

\title{First Observational Results from ALFA\\ with Natural and Laser
Guide Stars}
 
\author{R. I. Davies\supit{a}, W. Hackenberg\supit{a},
T. Ott\supit{a}, A. Eckart\supit{a}, S. Rabien\supit{a}, S. Anders\supit{a}
\skiplinehalf  
and
\skiplinehalf  
S. Hippler\supit{b}, M. Kaspar\supit{b}
\skiplinehalf  
\skiplinehalf  
\supit{a}Max-Planck-Institut f\"ur extraterrestrische Physik, 85740
Garching, Germany
\skiplinehalf  
\supit{b}Max-Planck-Institut f\"ur Astronomie, K\"onigstuhl 17, 69117
Heidelberg, Germany 
\skiplinehalf  
} 
 
\authorinfo{ \\
For further information contact R. Davies: davies@mpe.mpg.de\\
ALFA webpage: http://www.mpe.mpg.de/www\_ir/ALFA/ALFAindex.html}

\begin{document} 
  \maketitle  
 
\begin{abstract} 

The sodium laser guide star adaptive optics system ALFA, which is
installed at the Calar Alto 3.5-m telescope, has been undergoing an
intensive optimisation phase.
Observations using natural guide stars that are presented in this
paper, show that for bright stars (\mv$\simlt$8) 
it is now possible to reach K-band
Strehl ratios in excess of 60\% and to easily resolve binaries at the
diffraction limit of the telescope.
In more typical usage Strehl ratios in the range 10--15\% can be
achieved over a wide field ($\sim$2\arcmin); and the limiting magnitude
is currently \mv$\sim$12.5.
We also present some of the first
{\bf spectroscopy at diffraction limited scales},
showing we are able to distinguish spectra of binary stars with
a separation of only 0.26\arcsec.
Our last set of results are on Abell galaxy clusters, including a 
{\bf correction on a galaxy using the laser guide star} as the reference.
There are still a number of difficulties associated with the laser, 
but our best result to date is of the
galaxy UGC\,1347 in Abell\,262.
Correcting tip-tilt on a star 41\arcsec\ away and higher orders on the
laser, we achieved in increase in peak
intensity of 2.5, and a reduction in FWHM from 1.07\arcsec\ to
0.40\arcsec.
It is expected that further significant advances will be made in the
next 6 months and beyond.

\end{abstract} 
 
 
 
\section{INTRODUCTION} 
\label{sect:intro}  
 
During the last 6\,months significant progress has been made in the
quality of correction that can be achieved with ALFA, even for
mediocre atmospheric conditions.
This is a necessity at the Calar Alto observatory since the seeing is
often 1\arcsec\ or worse.
For a number of observing programmes, ALFA is now able to compete
effectively with other adaptive optics systems.

Some progress has also been made with the laser guide star, although
this is still limited by non-ideal observing conditions, being
particularly sensitive to atmospheric transmission.
We have been able to close the loop on the laser and correct the field
around the galaxy UGC\,1347 with an improvement in both peak intensity
and FWHM of a factor of 2.5.
There remain two main restrictions to regular observations with the
laser.
One of these is the beam jitter, which can often throw the LGS spots
outside the centroiding regions on the wavefront sensor.
The second is the LGS size ($\sim$2.5\arcsec), the cause of which we are
currently investigating;
an intended fibre link to replace the beam relay between the laser and
the telescope may overcome this.
As well as these we are implementing automatic control algorithms for
laser tuning and focussing, and also WFS focussing.
These should increase the observing efficiency by a large margin.

In this contribution we present the observational results that have
been obtained so far.
We begin by noting in Section~\ref{sec:best} the best performance that
ALFA has achieved.
In Section~\ref{sec:3d} the first spectroscopy at diffraction limited
scales is presented.
Section~\ref{sec:bd40} continues with observations of an
$80\times80$\arcsec$^2$ field around the Herbig Ae/Be star, to look at
how the correction varies at large distances off-axis.
Results from an imaging study of Abell cluster galaxies are given in
Section~\ref{sec:abell}. 
We concentrate in particular on two galaxies, correcting with a nearby
natural guide star on one, and using the laser guide star as a
wavefront reference for the other.

\section{Best Performance}
\label{sec:best}

During 1998 the ALFA adaptive optics hardware, software, and personnel
underwent intensive optimisation and the system is now beginning to
deliver the specified performances.
These criteria included achieving 40\% Strehl in the K-band with median
0.9\arcsec\ seeing.

For the purposes of this contribution, 
Strehl ratios have been calculated from the ratio of peak to total
fluxes in the PSF image, by comparing it to that calculated for the
theoretical diffraction limited PSF (with a 3.5\,m mirror and
1.37\,m central obscuration).
Unless stated, they take no account of where in the pixel the PSF is
centred and hence may underestimate the actual Strehl.
This error is rather variable: for example with 0.04\arcsec\
pixels and a high Strehl measurement might be $50\pm5$\%, or with
0.08\arcsec\ pixels a lower Strehl measurement could be $15\pm5$\%.

For seeing of around 1\arcsec, K-band Strehls in excess of
60\% can be reached for the brightest stars (\mv$\simlt$6), while
values in the range 25--50\% can be attained for stars with
\mv$\simlt$8.
These results are very impressive considering that even if 32 (Zernike)
modes are corrected {\em perfectly}, the residual wavefront error is 
$\sigma^2 = 0.43$\,rad$^2$ giving a maximum theoretical Strehl ratio of
only 65\%; and this does not include bandwidth limitations, noise, or
mirror flatness and other residual static aberrations.
The performance can be translated to other wavebands, and we have
achieved a J-band Strehl of 12\% on SAO\,56114 (\mv=7.0), and a
resolution better than 0.10\arcsec, close to the diffraction limit of
0.07\arcsec\ FWHM.

\begin{figure}[ht]
\plotfiddle{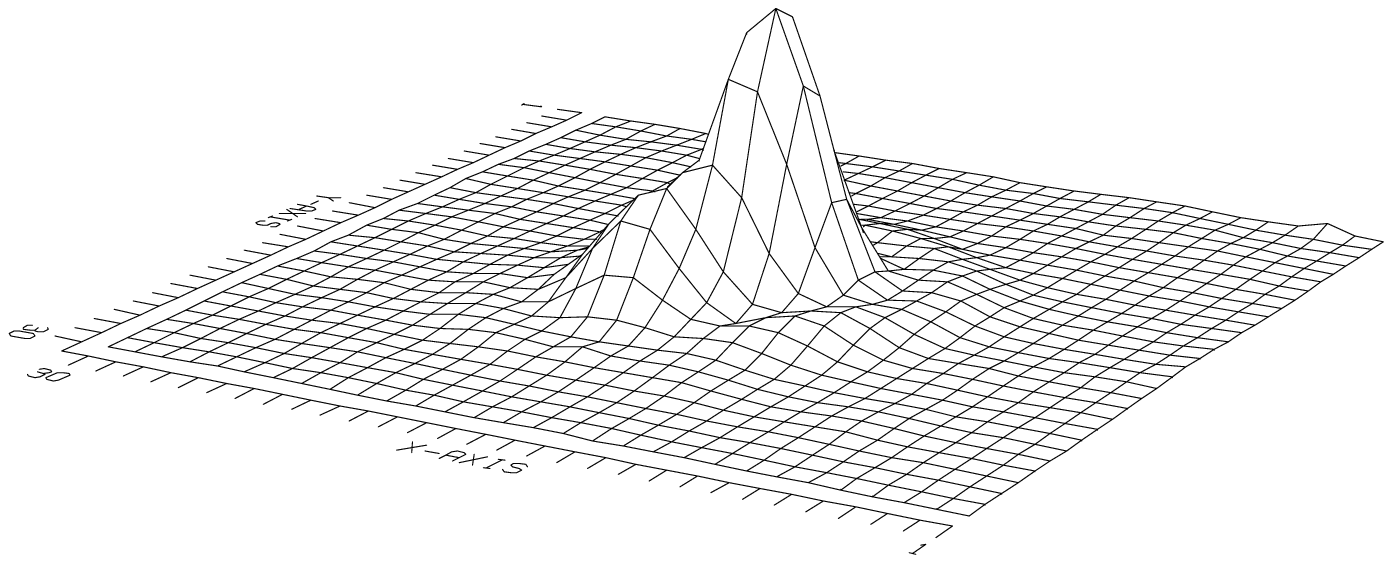}{4cm}{0}{60}{60}{-300}{-200}
\plotfiddle{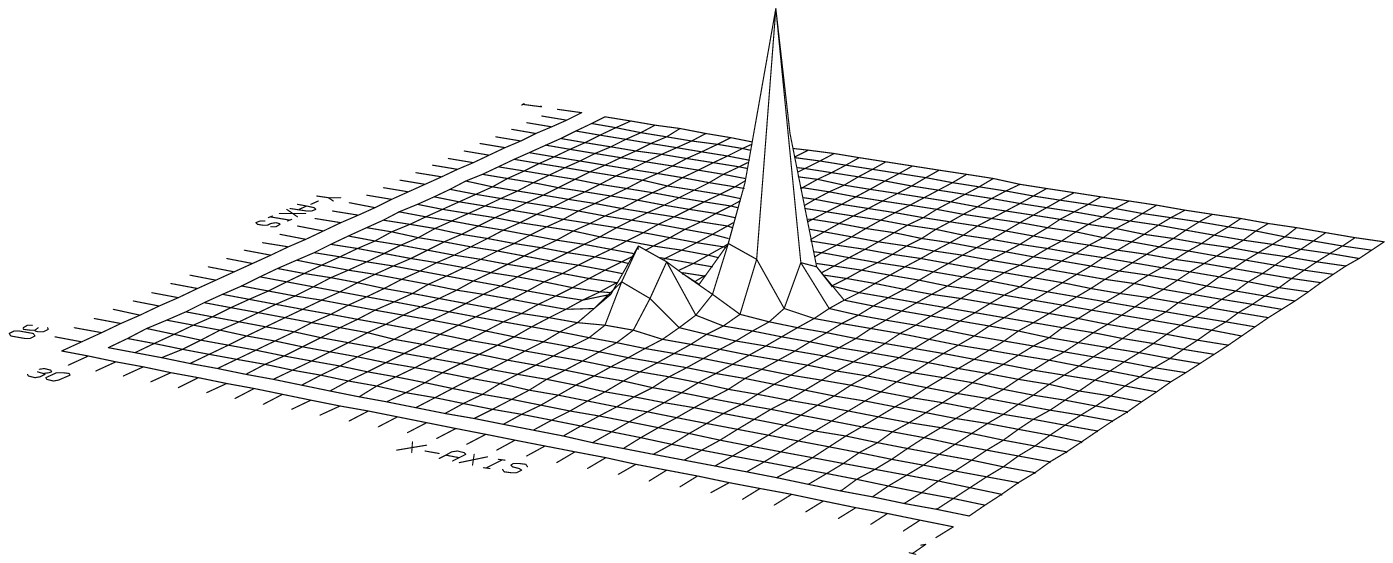}{0cm}{0}{60}{60}{-60}{-175}
\caption{K-band surface plots of SAO\,36874 (\mv=4.9), observed with a
pixel scale of 0.04\arcsec, and correcting 32
modes at a frame rate of 200\,Hz.
Left: a direct image.
Right: a deconvolved image (30 iterations of the Lucy algorithm);
the PSF star (SAO\,56114, \mv=7.0) had a Strehl ratio of 45\% and a
FWHM of 0.14\arcsec. 
The two stars of SAO\,36874 are clearly resolved at a separation of
0.15\arcsec, only marginally greater than the K-band
diffraction limit of the telescope.
\label{fig:double}}
\end{figure}

Any observation with a Strehl ratio greater than about 20\% will
effectively be diffraction limited.
An excellent example of this is the serendipitous discovery of a double
star in SAO\,36784 as shown in Fig.~\ref{fig:double}, during testing of
control parameters.
This star is listed in the Washington Double Star Catalog (WDS, Worley \&
Douglass 1997)\nocite{wor97} as the
primary partner of a pair with separation of 20.5$''$ and magnitudes
\mv=6.0 and 12.3;
ALFA has now shown that the primary itself is double, although it is
unclear whether this is a true binary system or a projection effect.
The separation of the two components is 0.15\arcsec, almost at the
telescope's diffraction limit (0.13\arcsec\ FWHM), and they can clearly
be discriminated after deconvolution.
The observation shows the vast potential for studies of
stars in multiple systems, for determining orbits and system stability;
as well as for characterising spectroscopic and speckle binaries, by
measuring broad-band colours or individual spectra.
An example of the latter is given in next Section~\ref{sec:3d}.

\section{Integral Field Spectroscopy: HEI\,7}
\label{sec:3d}

The potential for spectroscopy at diffraction limited scales
represents an exciting aspect of adaptive optics.
However, the inherent difficulties also make it a special challenge.
Standard techniques with a longslit may no longer be feasible, since
in order to make the most of the benefits of minimising both background
and contamination from extended sources, the slit must be extremely narrow.
Accurately positioning a target on such a slit can be very time
consuming, and for faint sources this becomes impractical.
The alternative method of integral field spectroscopy which involves
re-ordering a 2-dimensional field onto a longslit, dispersing it, and
reconstructing a datacube (2 spatial and 1 spectral axes) afterwards,
opens considerable opportunities in this area.

\begin{figure}
\plotfiddle{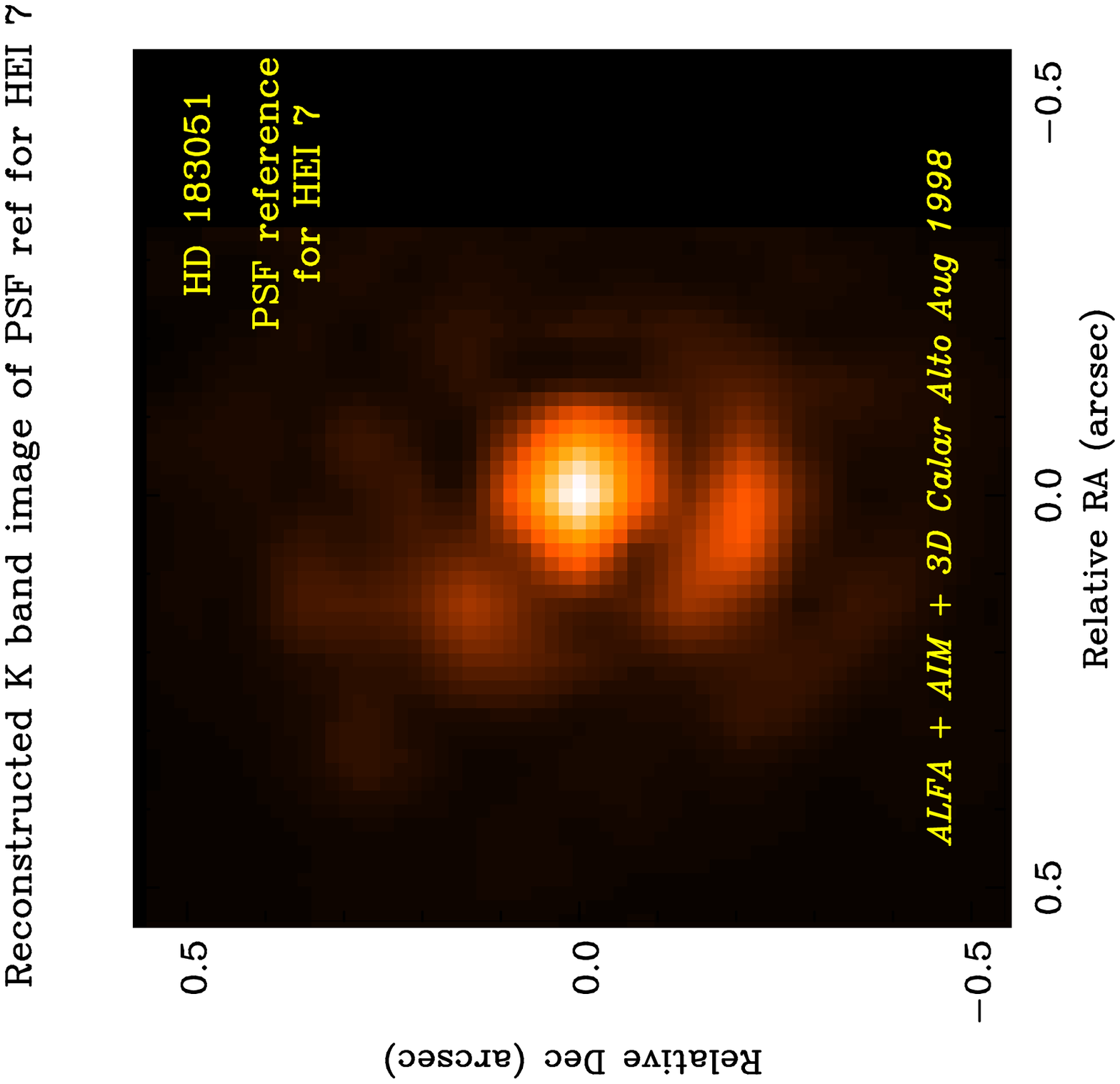}{6cm}{-90}{40}{40}{-280}{200}
\plotfiddle{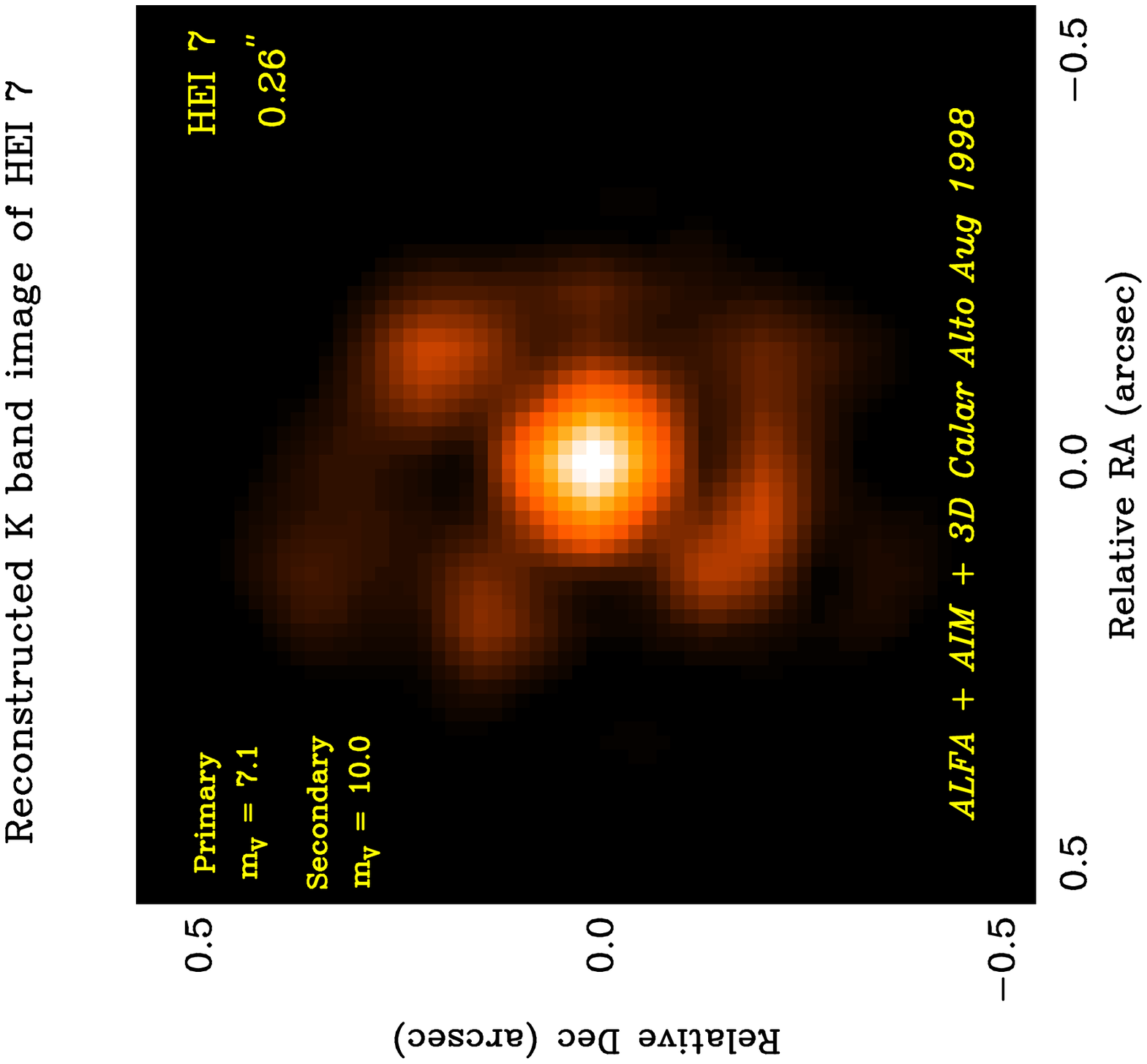}{0cm}{-90}{40}{40}{-50}{224}

\caption{Reconstructed K-band images of the PSF reference HD\,183051
(left) and the binary HEI\,7 (right), rebinned to a smaller pixel scale
(original scale gives $16\times16$ pixels).
North and East are rotated 15\deg\ from down and right respectively.
Parts of the first diffraction ring can be seen;
the extra blob to the top right in HEI\,7 is the companion.
\label{fig:hei7.images}}
\end{figure}

\begin{figure}
\plotfiddle{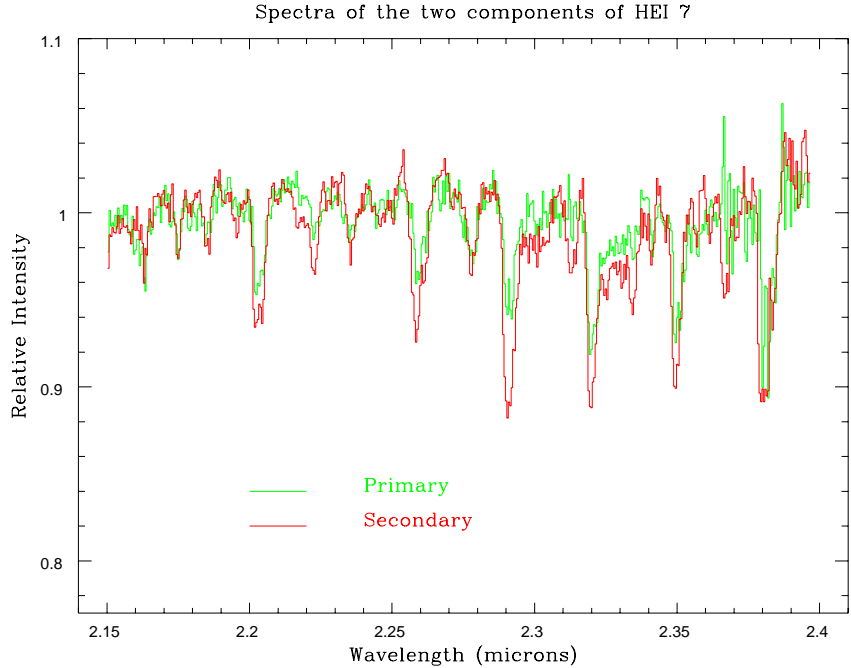}{9cm}{-90}{45}{45}{-180}{260}
\caption{Smoothed K-band spectra of the two components of HEI\,7.
All the features show the expected depth, based on spectral type
classifications (K0\,V and M0\,V) of the two stars.
In particular the $^{12}$CO bandheads at 2.29, 2.32, 2.35, and
2.38\um\ vary by almost a factor of 2, while the Mg line at 2.28\um\ is
the same for the two types.
Other prominent lines include Na (2.21\um) and Ca (2.26\um).
\label{fig:hei7.spectra}}
\end{figure}

During August 1998, the MPE 3D imaging spectrometer (Weitzel et
al. 1996)\nocite{wei96}, which obtains simultaneously H- or K-band
spectra of an entire $16\times16$ pixel field, was used with ALFA.
In order to facilitate this, an aperture interchange module ({\it AIM},
Anders et al. 1998a\nocite{and98a}) has been
built as an interface to allow the pixel scale to be changed between
0.25$''$ and 0.07$''$ per pixel, as well as providing the ability for
efficient sky observations with minimal overhead.
Although the field of view is small (1.2$''$), this is ideally matched
for observations of AGN or close binaries.
Here we present some of the first spectroscopy at diffraction limited
scales (see Anders et al. 1998b)\nocite{and98b}, of HEI\,7, a
binary listed in the WDS Catalog.
The primary, more commonly known as HD\,197443, is itself a
photometric/spectroscopic binary (denoted AB) with a period of about
6\,hrs and has not been resolved.
A third member of the system (denoted C) was suspected from variations
in the time of 
minimum in the primary pair, and its orbit was first calculated by 
Hershey (1975)\nocite{her75} from changes in parallax of the primary
with respect to a set of reference stars (later confirmed by observation).
The data yielded both the parameters of the AB-C 30.5\,yr orbit as well
as the absolute parallax, which sets the distance as 24\,pc (ie the
current projected separation is only 6\.AU).

A PSF reference (HD\,183051, left in Fig.~\ref{fig:hei7.images}) with the
same \mv=7.1 as the primary component of HEI\,7 was observed 10 minutes
before the binary using the same adaptive optics parameters.
Although the first diffraction ring was complete when calibrating on
the fibre at the beginning of the night, it is much more patchy in the
stellar image.
Even so, the difference between this and the reconstructed K-band image of
HEI\,7 is clear: the secondary can be seen to the top right (south
east) at a radial separation of 0.26$''$ and about a factor 15 fainter
(\mv=10.0).

Spectra of the two components are shown in
Fig.~\ref{fig:hei7.spectra}.
The primary (AB) component has the expected absorption features for a
K0\,V star;
the differences between this and the secondary (C) are expected from its
spectral type of M0\,V (estimated from colours).
In particular the $^{12}$CO bandheads are almost a factor of 2 deeper,
while the Mg line has a similar equivalent width.
The depth of the Na (2.21\um) and Ca (2.26\um) lines also shows some
variation.
These spectra, then, are from two separate objects and not
simply the same one extracted at different points.
This example clearly demonstrates the feasibility
of diffraction limited spectroscopy, although the technique remains
difficult.
It is fortuitous that the diffraction ring is faint in the region near
the secondary star so that deconvolution is unnecessary.
The next step is to develop methods which will allow overlapping
spectra to be distinguished.

\section{Wide Field Correction: BD\,+40\deg\,4124}
\label{sec:bd40}

\begin{figure}
\plotfiddle{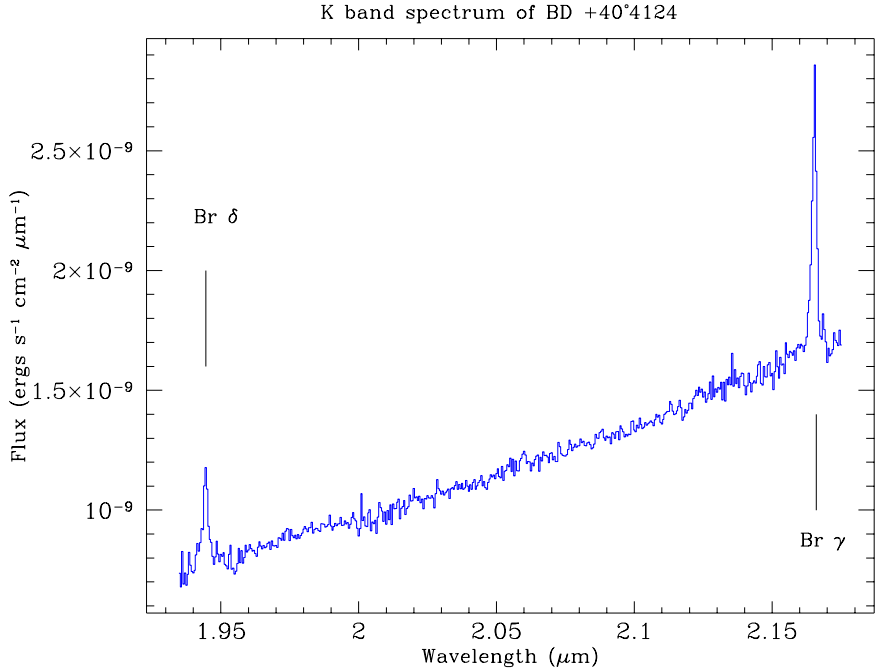}{8.5cm}{-90}{45}{45}{-180}{250}
\caption{K-band specturm of BD\,+40\deg\,4124.
The continuum appears to be featureless axcept for the Brackett lines
at 1.95 \& 2.17\um, suggesting an extinction of \av=29\,mag.
\label{fig:bd40.spectrum}}
\end{figure}

\begin{figure}
\plotfiddle{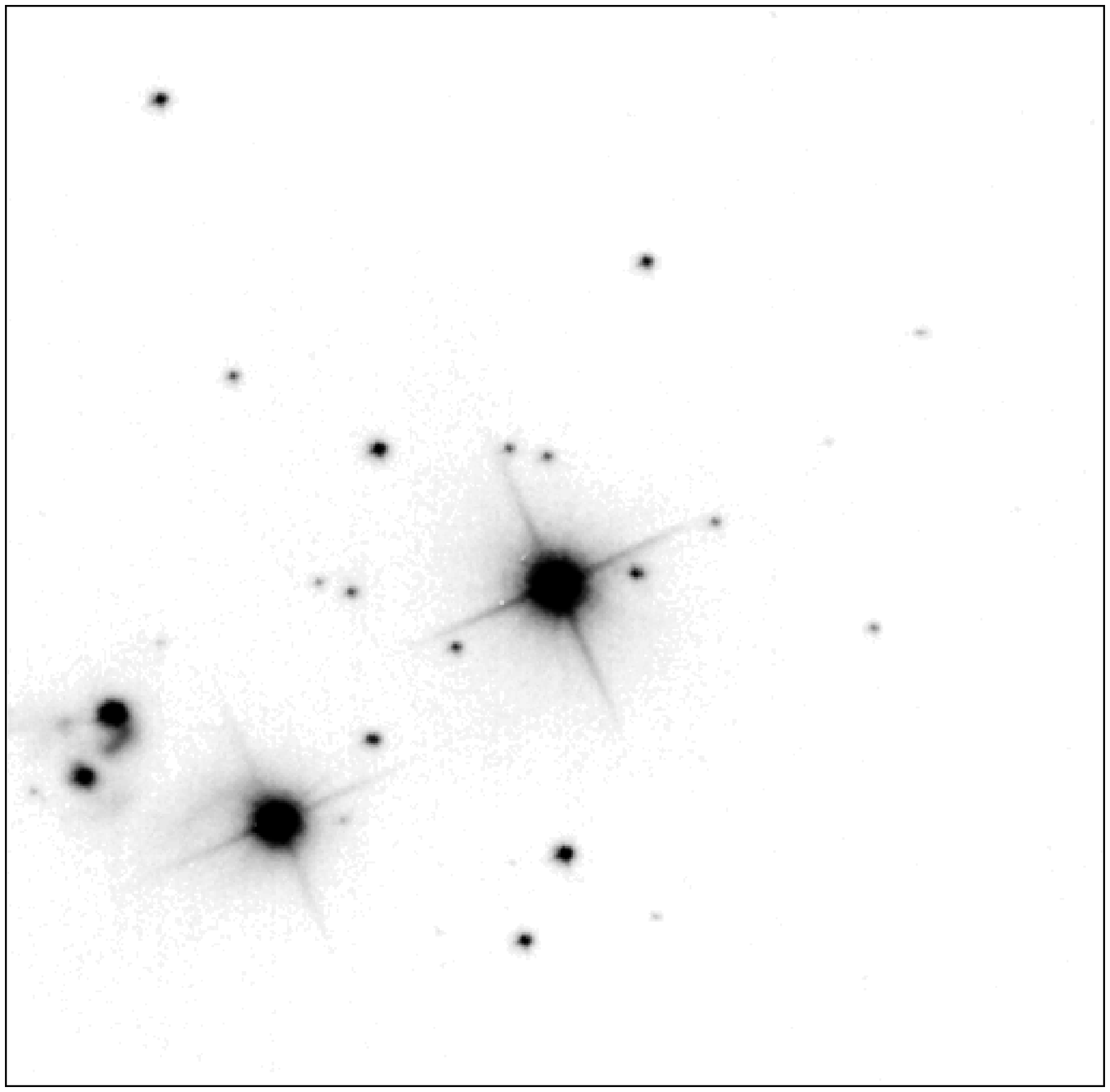}{9.5cm}{0}{60}{60}{-170}{-130}
\plotfiddle{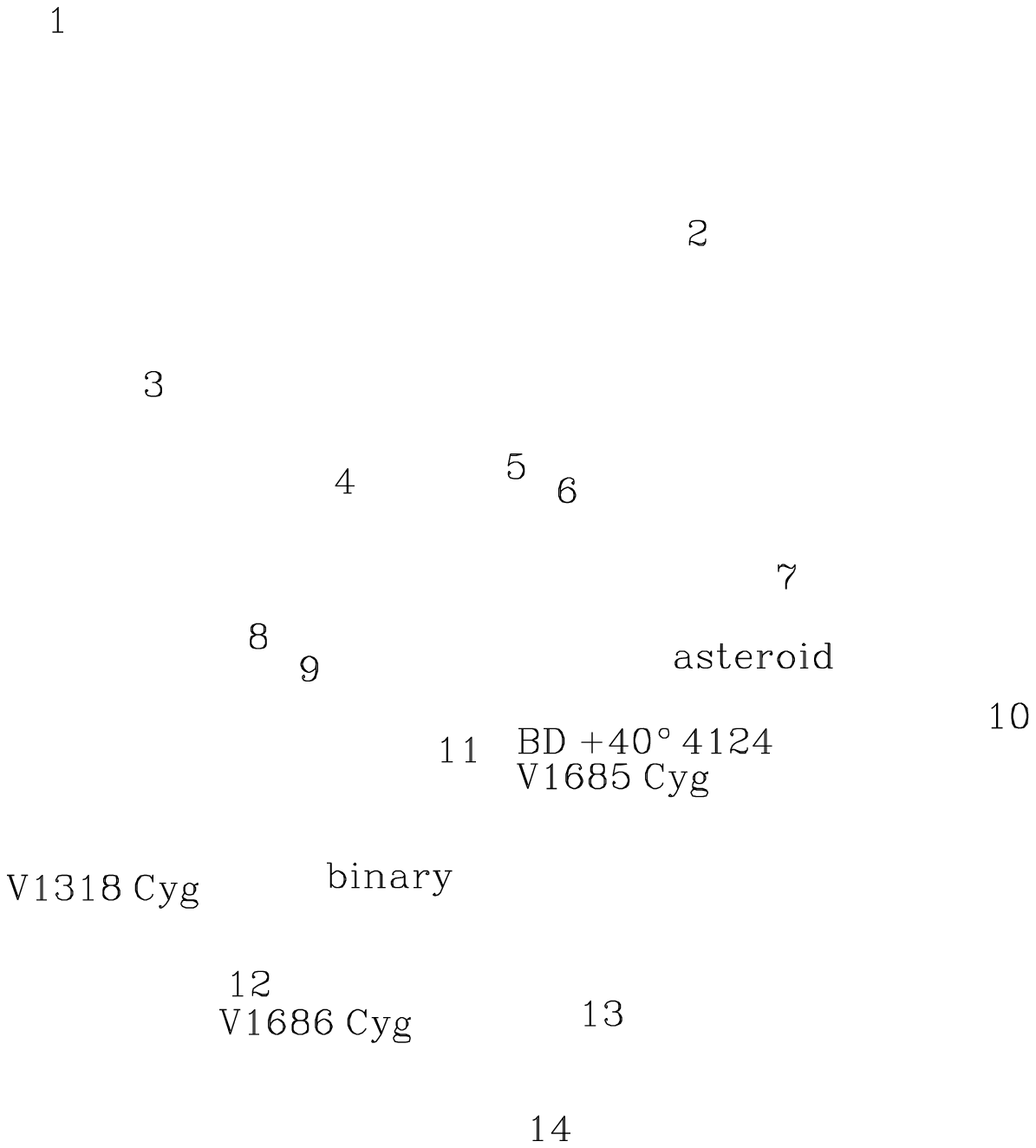}{0cm}{0}{63}{63}{-190}{-136}
\caption{A K-band image of the $80'' \times 80''$ field around the
 Herbig Ae/Be star BD\,+40\deg\,4124.
North and East are 22\deg\ clockwise from up and left, respectively.
It was used for the AO, but was only bright enough
 (\mv=10.6) to correct
18 modes at a frame rate of 75\,Hz.
Strehl ratios of 10--15\% and resolutions better than 0.22$''$ 
were obtained.
The effects of anisoplanaticism are shown in the next figure.}
\label{fig:bd40.field}
\end{figure}

The 3D spectrometer was also used to obtain a K-band spectrum of the
Herbig Ae/Be star BD\,+40\deg\,4124 (\mv=10.6, \mk=5.6)
(Fig.~\ref{fig:bd40.spectrum}), 
while the field around this star was imaged in the JHK bands using
$\Omega$-Cass (Fig.~\ref{fig:bd40.field}).

Herbig Ae/Be stars are the massive counterparts ($\simgt$3\msun) to
the low mass T~Tauri pre-main-sequence stars, typically being deeply
embedded in gas and dust, but with strong hydrogen recombination
lines.
BD\,+40\deg\,4124 lies in a small aggregate containing a few other similar
stars (the nearest are another Herbig Ae/Be star V1686\,Cyg, and the
pair V1318\,Cyg), but which is
isolated from any large star-forming complex (Hillenbrand et
al. 1995)\nocite{hil95}.
Since it is by far the most massive (see Table~\ref{tab:bd40}) its
effects on the local environment can be studied independently, and
at a distance of marginally less than 1000\,pc (1$'' \equiv 1000$\,AU)
adaptive optics can provide the resolution on solar-system scales which
is required for studying the dust envelopes around these stars.
As all the stars in the group have ages in the range
$10^5$--$10^6$\,yrs, this can be of some importance in understanding
how massive stars are formed; and whether the winds, outflows, and
radiation field around them influence the evolution of the aggregate.

\begin{table}[ht]
\centering
\caption{BD\,+40\deg\,4124 region: stellar parameters\label{tab:bd40}}
\smallskip
\begin{tabular}{llccc}
\hline
Star & Spectral Type & \av /mag & log\,L/\lsun & M/\msun \smallskip \\
\hline \hline
BD\,+40\deg\,4124 & B2 Ve     & 3.6--3.8 & 4.10 & 13   \\
V1686\,Cyg        & B5 Ve     & 4.2--6.7 & 2.77 & 4.5  \\
V1318\,Cyg North  & mid A--Fe & 7--15    & --   & $>1$ \\
V1318\,Cyg South  & mid A--Fe & 8--15    & --   & $>1$ \\
\hline
\end{tabular}

\smallskip Note: all data from Hillenbrand et al. (1995)
\end{table}

The spectrum of BD\,+40\deg\,4124 in Fig.~\ref{fig:bd40.spectrum} is
remarkable in its lack of any features other than Brackett lines.
The ratio of these lines \brg/Br$\delta = 2.83$ suggests an extinction
of \av=29\,mag assuming an intrinsic ratio of 1.52 (Osterbrock
1989)\nocite{ost89} and either the A$_\lambda \propto
\lambda^{-1.85}$ extinction law of Landini (1984)\nocite{lan84} or the
curves from Howarth (1983)\nocite{how83}.
This is rather higher than that given in
Table~\ref{tab:bd40} derived from optical lines, but is consistent with
the \av=45\,mag estimated via $N_{\rm H}$ from $^{12}$CO and $^{13}$CO
luminosities (Hillenbrand et al. 1995).
These observations can be reconciled if the line-emitting gas is mixed
with the dust obscuring it, so that the optical lines sample only the
surface regions and underestimate the extinction-corrected flux.
However, the dust must then be internal to the \hiiregion\ 
surrounding the star, and as much as 50--90\% of the Lyman
continuum photons would be absorbed before they could be processed into
recombination lines (Wood \& Chruchwell 1989)\nocite{woo89} reducing
the recombination luminosity at all wavelengths.
An alternative scenario is that the dust exhibits a dense clumpy
structure so that the optical lines can sample the \hiiregion\ through
only the least-obscured lines-of-sight, the near-infrared lines denser
regions, and the radio lines can probe through even the densest
clumps.

\begin{figure}
\plotfiddle{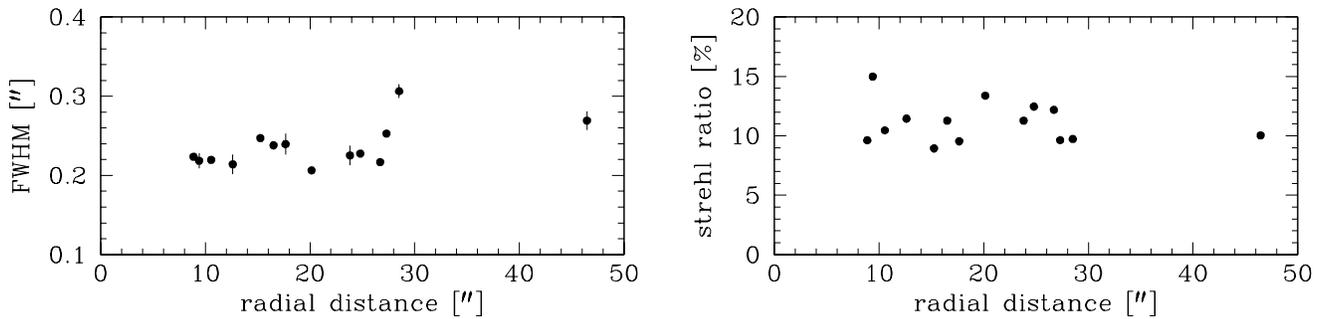}{4cm}{-90}{70}{70}{-277}{400}
\caption{Strehl ratio and FWHM of 14 stars in the field around
BD\,+40\deg\,4124, plotted against distance from this star.
The advantage of working in the partial correction regime is clear,
since the effects of anisoplanaticism are much reduced:
only beyond 30$''$ is the a small reduction in performance.}
\label{fig:bd40.strehl-fwhm}
\end{figure}

The wide field K-band image taken with $\Omega$-Cass at a pixel scale of
0.08\arcsec\ is shown in Fig.~\ref{fig:bd40.field}.
BD\,+40\deg\,4124 was centered in the wavefront sensor,
allowing correction of 18 modes with a frame rate of 75\,Hz.
Although this star was saturated the Strehl ratios and FWHMs of 14
others in the $80''\times80''$ field were measured, and these are
plotted against their radial 
distance in Figure~\ref{fig:bd40.strehl-fwhm}.
The FWHMs were estimated using only vertical and horizontal cuts (the
`errorbars' are lines joining these points) since models suggest that
the pixel size is already limiting the resolution:
for Strehls of 10--15\% an intrinsic resolution of 0.16\arcsec\ is
expected, but the measurable resolution is 0.2--0.3\arcsec\ if the flux
is binned into 0.08\arcsec\ pixels.
The figure indicates that out to radii of at least 30$''$ there is very
little degradation in performance, since the wavefront error due to
anisoplanaticism is small compared to the total error;
and beyond this as far as can be measured, the correction is only
marginally worse.

\begin{figure}
\plotfiddle{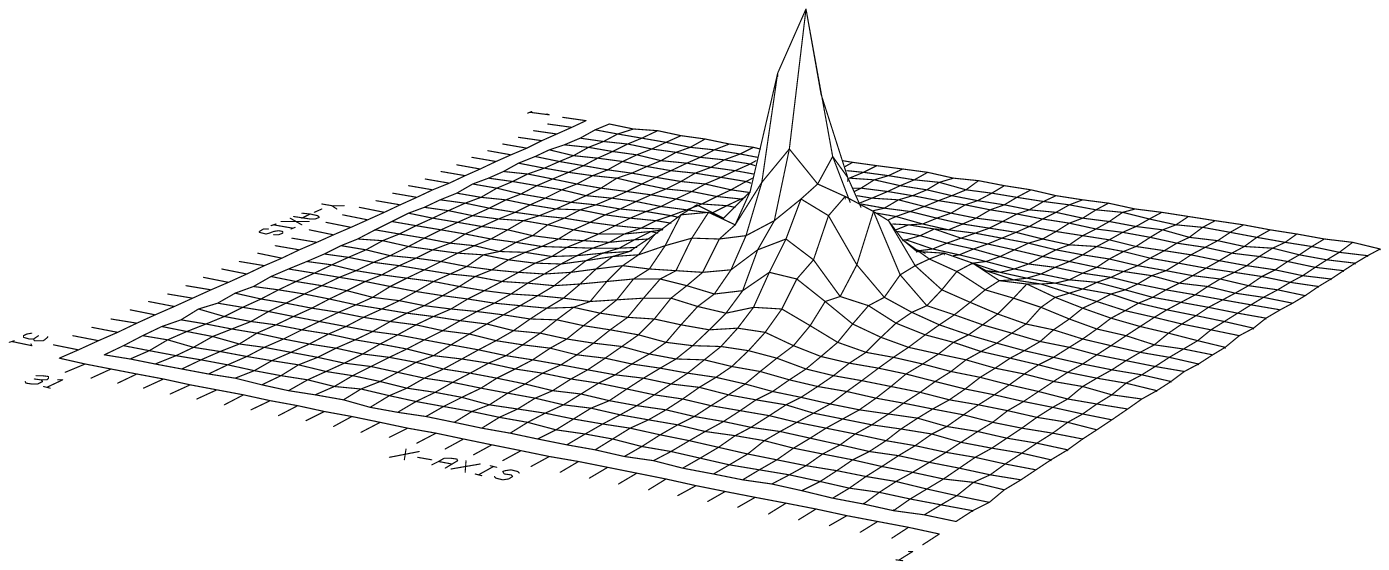}{4cm}{0}{60}{60}{-300}{-200}
\plotfiddle{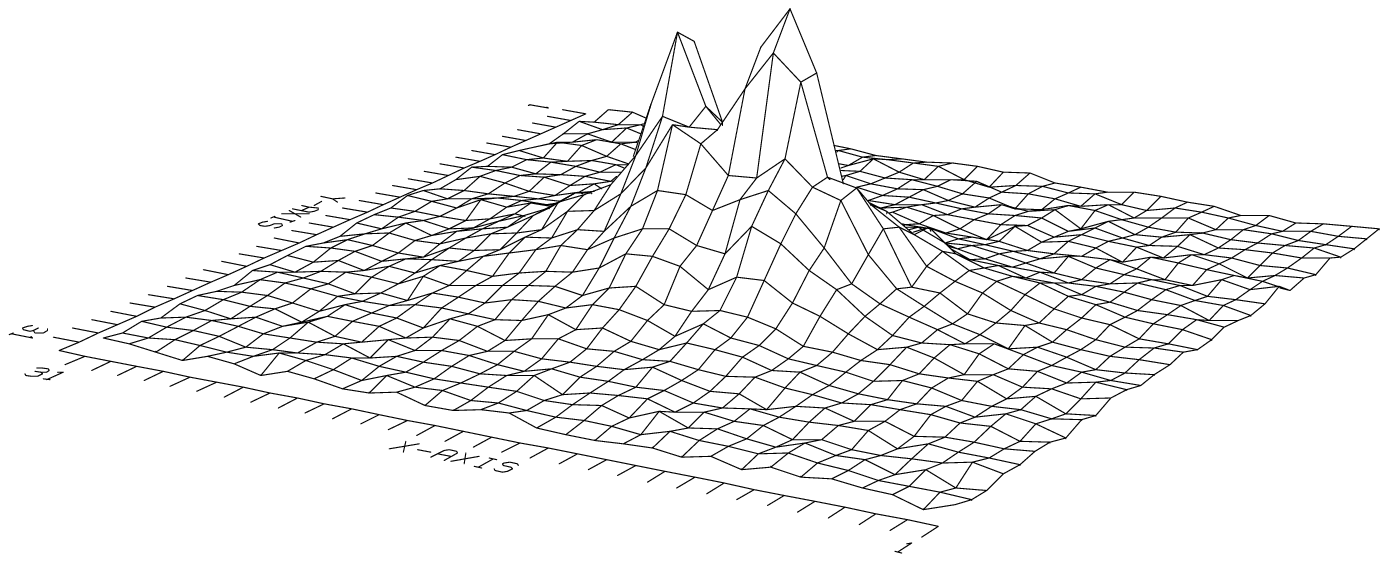}{0cm}{0}{60}{60}{-60}{-175}
 \caption{K-band surface plots of star 14 (left) and the binary system
 (right), both approximately 20$''$ from BD\,+40\deg\,4124.
The pixels are 0.08$''$ across and the resolution is
 sufficient to easily separate stars 0.32$''$
 apart, as shown by the previously unknown binary system here.
The profiles of all the stars show that the PSF changes very little
 even out to distances of 50$''$.}
 \label{fig:bd40.psf}
\end{figure}

Figure~\ref{fig:bd40.psf} shows that the correction achieved here is
useful: a star 20$''$ away clearly has a sharp peak (as do all the
single stars in the field), while another has
a double peak, identifying it unambiguously as a previously unknown
binary system with a separation of 0.32$''$.
Previous studies of multiplicity in Herbig Ae/Be stars (eg Leinert et
al. 1997)\nocite{lein97} have had to resort to speckle imaging, a slow
and limiting process compared to adaptive optics.
Importantly, examination of profiles of all the sources indicates
that the speckle pattern in the halo appears to be stable over the
whole field.
This means that deconvolution should work well (yielding an effective
resolution rather better than 0.2$''$) and that the choice of
PSF, at least in this case, is not critical.
A further result from this image was that the southern star of the
V1318\,Cyg pair was resolved to be
$0.4''\times0.3''$, without the expected sharp core.
The extinction to this object derived
from the 1-0\,Q(3)/1-0\,S(1) ratio (Aspin et al. 1994)\nocite{asp94} is
\av=45$\pm$20\,mag, using the same laws as above.
Combined with very red JHK colours ($J-K = 4$--5\,mag) this has led
to interpretation as a dense circumstellar dust shell.
We estimate the deconvolved size to be $300\times170$\,AU.
Assuming the normal gas/dust ratio of 100, this is equivalent to
$N_{\rm H} = 7\times10^{22}$\pcmsq, and a gas density on the order of
$10^6$\pcmcu.
At such densities vibrational levels of \h2\ above $\nu = 1$ are
collisionally de-excited, leading to large ratios
between the $\nu = 1$ and higher levels whatever the excitation
mechanism, for example \s1/2-1\,S(1)$>$10 (Aspin et al. 1994).

\section{Natural and Laser Guide Star Observations\\ of Cluster Galaxies}
\label{sec:abell}

Between December~1997 and August~1998 we observed 25 galaxies in
the Abell\,1367 and Abell\,262 clusters, including two barred spiral
galaxies UGC\,1344 and UGC\,1347.

\subsection*{UGC\,1344 with a Natural Guide Star}

\begin{figure}
\plotfiddle{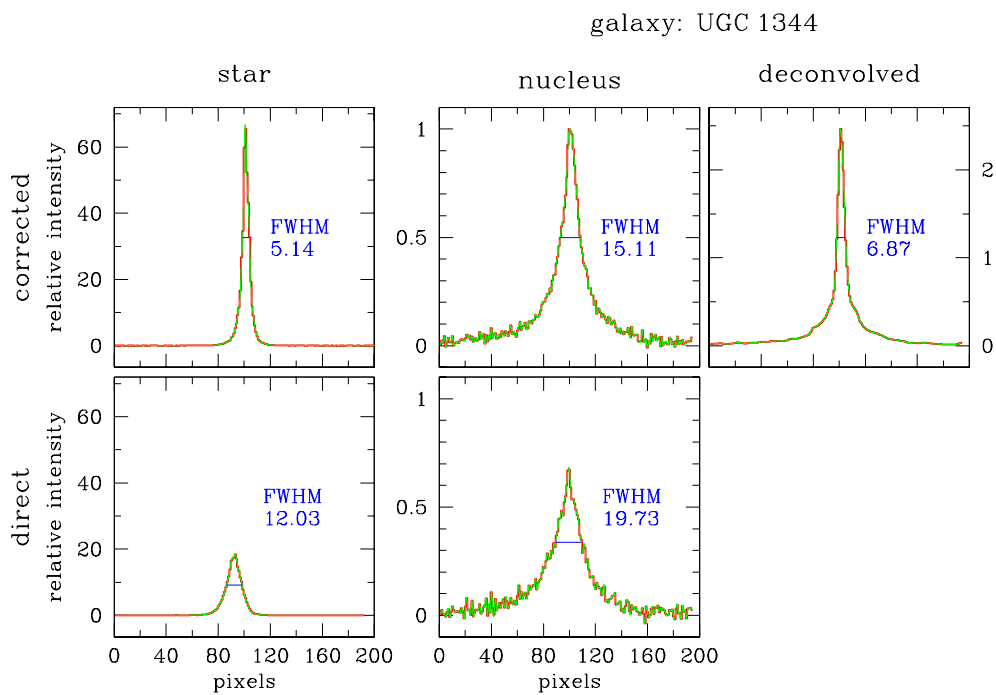}{9.5cm}{0}{70}{70}{-215}{-120}
\caption{Profiles of the star (left) and nucleus of UGC\,1344 (right)
 in both open loop (lower) and closed loop (upper);
a deconvolved closed loop profile is also shown (far right).
FWHM are given in pixels, with a scale of 0.08\arcsec.
The peak intensity of the star increases by a factor of 3, while the
galaxy nucleus increases by a factor of only 1.4 suggesting that there
is a narrow core component and a wider bulge component.
\label{fig:ugc1344}}
\end{figure}

For UGC\,1344 we used a nearby (27$''$) natural guide star with
V=11.0\,mag, correcting 7 modes at a sampling rate of 150\,Hz,
achieving a disturbance rejection bandwidth of about 13\,Hz.
Cuts of the profiles through the star and galaxy, for both open and
closed loop as well as deconvolved are shown in
Figure~\ref{fig:ugc1344}.
The peak intensity of the star increased by a factor of 3, and the FWHM
improved from 1.0$''$ to 0.4$''$.
For the galaxy, the peak increased by only 1.4, although
the discussion for BD\,+40\deg\,4124 indicates that the galaxy is
well within the isoplanatic patch.
The deconvolved image begins to indicate the reason: there is a narrow
only marginally resolved core, and a wider bulge component so the 
increase in peak intensity will effectively be that due to the core only.
The observations can be explained if the fluxes in the bulge and core
are similar.
The presence of a core suggests a recent localised burst of star
formation in the galaxy's nucleus.

\begin{figure}
\epsscale{0.4}
\plotone{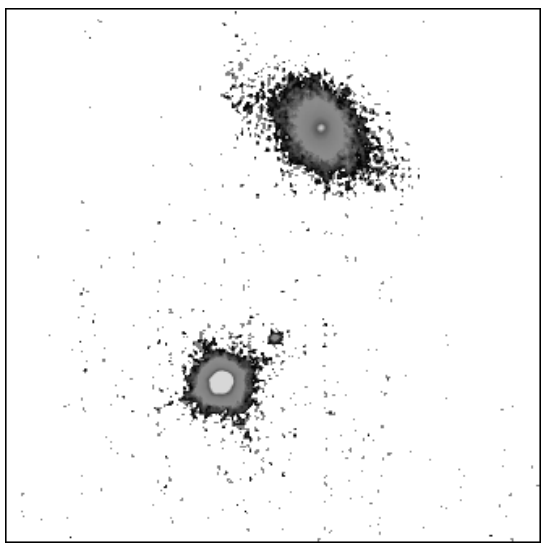}
\caption{Closed loop K-band image of UGC\,1344 and below it the
adaptive optics reference star (\mv=11.0 allowing correction of 18
modes with a frame rate of 75\,Hz).
The scale is given by their separation of 27\arcsec;
North and East are 22\deg\ clockwise from up and left respectively.}
\label{fig:u1344.image}
\end{figure}

\subsection*{UGC\,1347 with the Laser Guide Star}

\begin{figure}
\epsscale{0.4}
\plotone{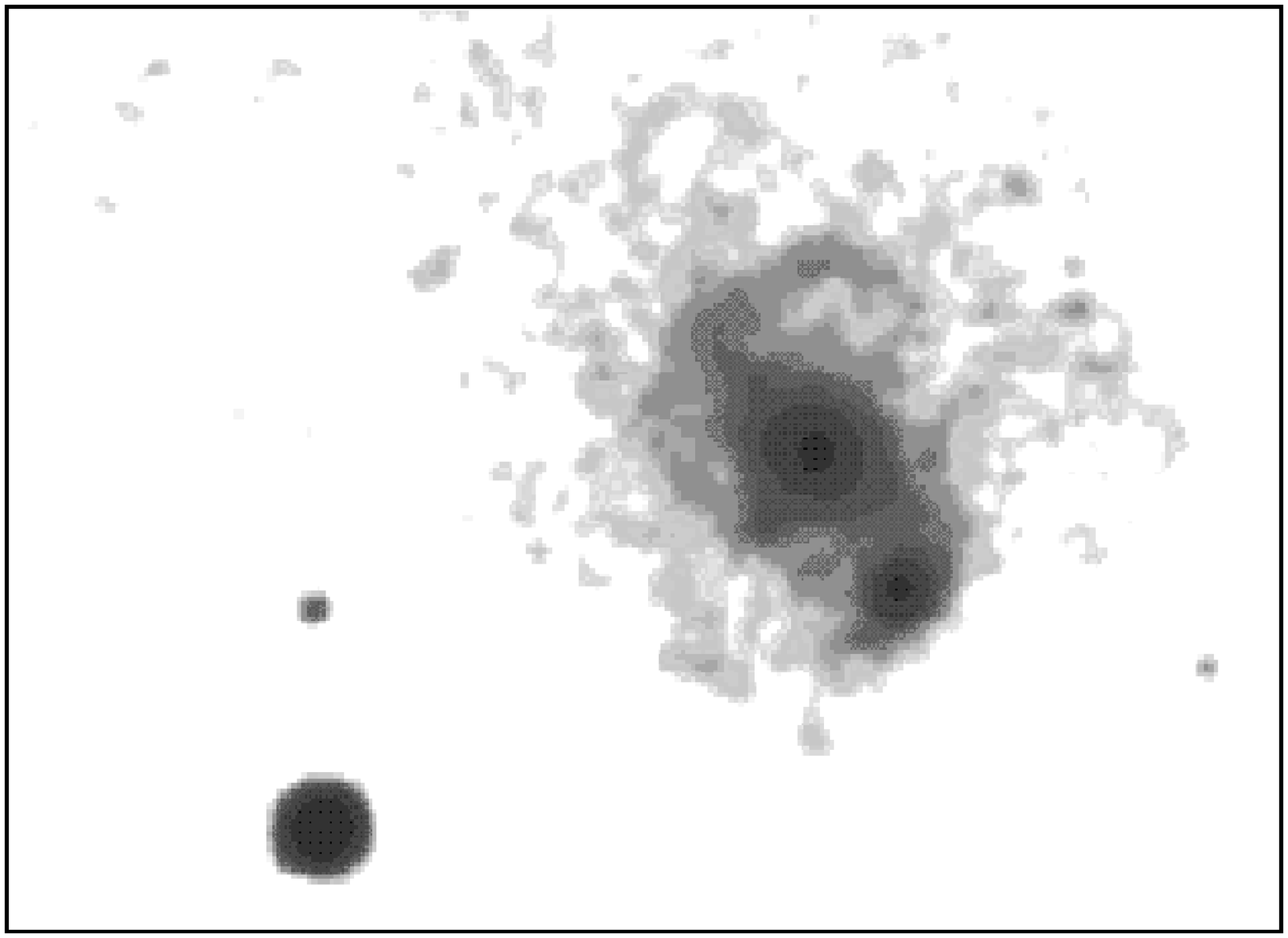}
\caption{Open loop K-band image of UGC\,1347 (nucleus, and below it a
compact \hiiregion) and to the South-East the star used as the tip-tilt
reference, smoothed to highlight the extended structure.
The scale is given by their separation of the star from the galaxy
nucleus of 41\arcsec;
North and East are up and left respectively.}
\label{fig:u1347.image}
\end{figure}

\begin{figure}
\epsscale{.2}
\plotone{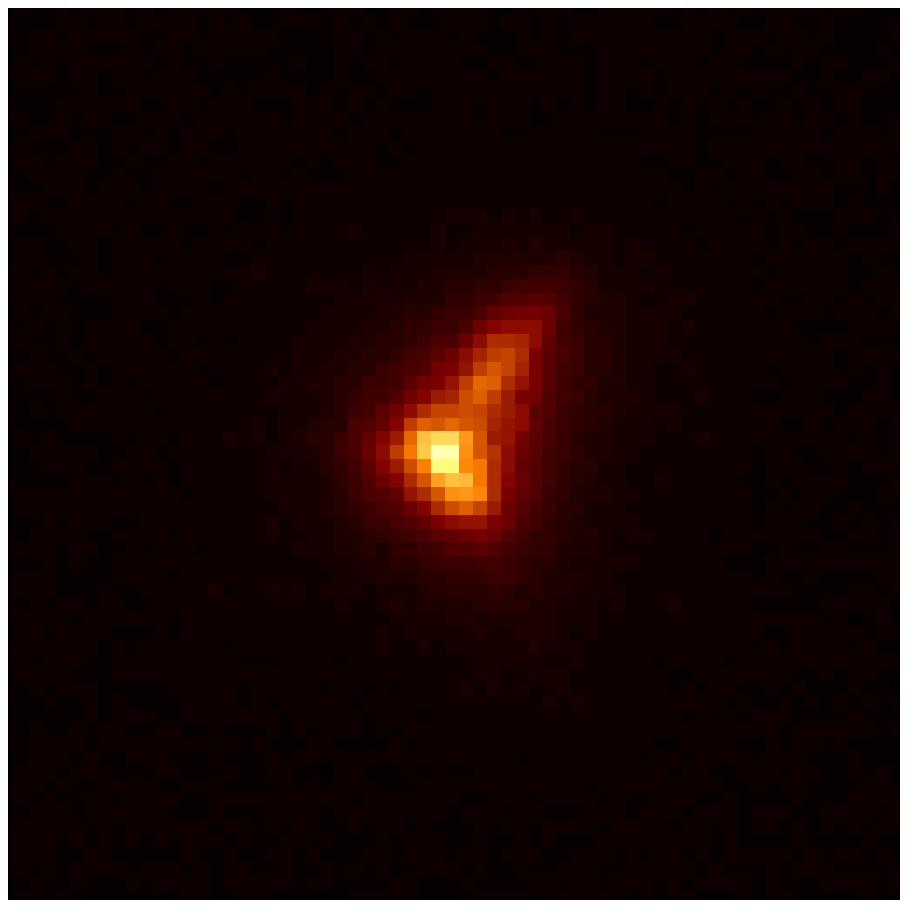}
\hspace{8mm}
\plotone{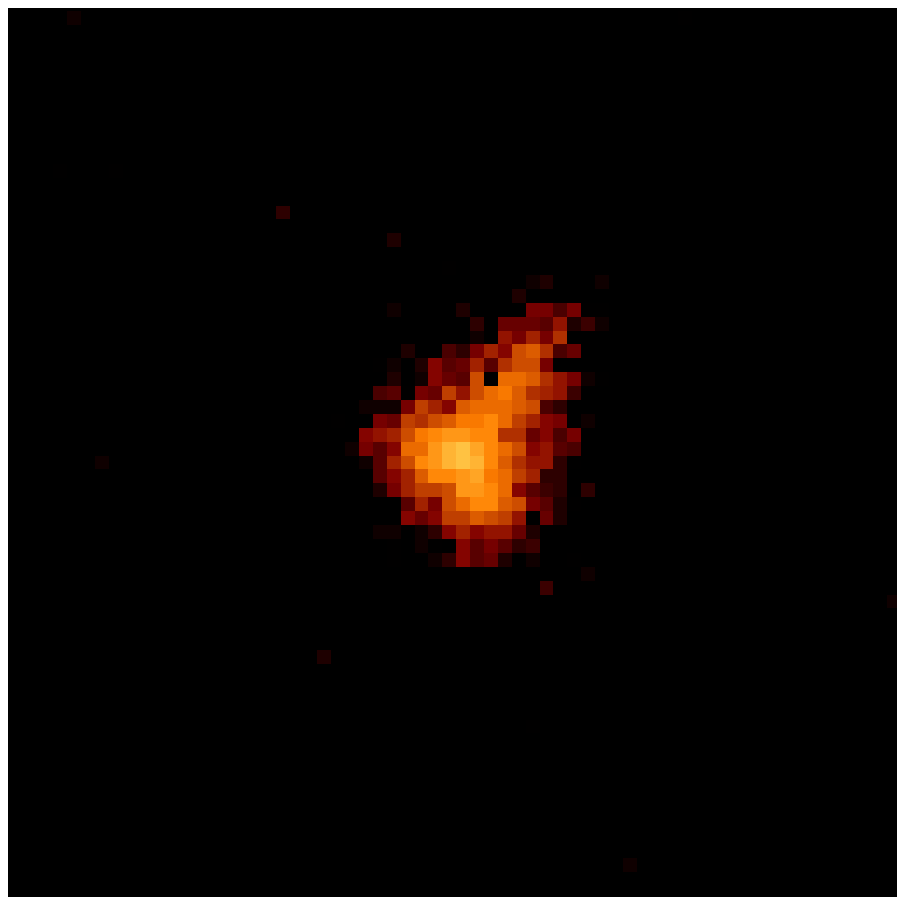}
\hspace{1mm}
\plotone{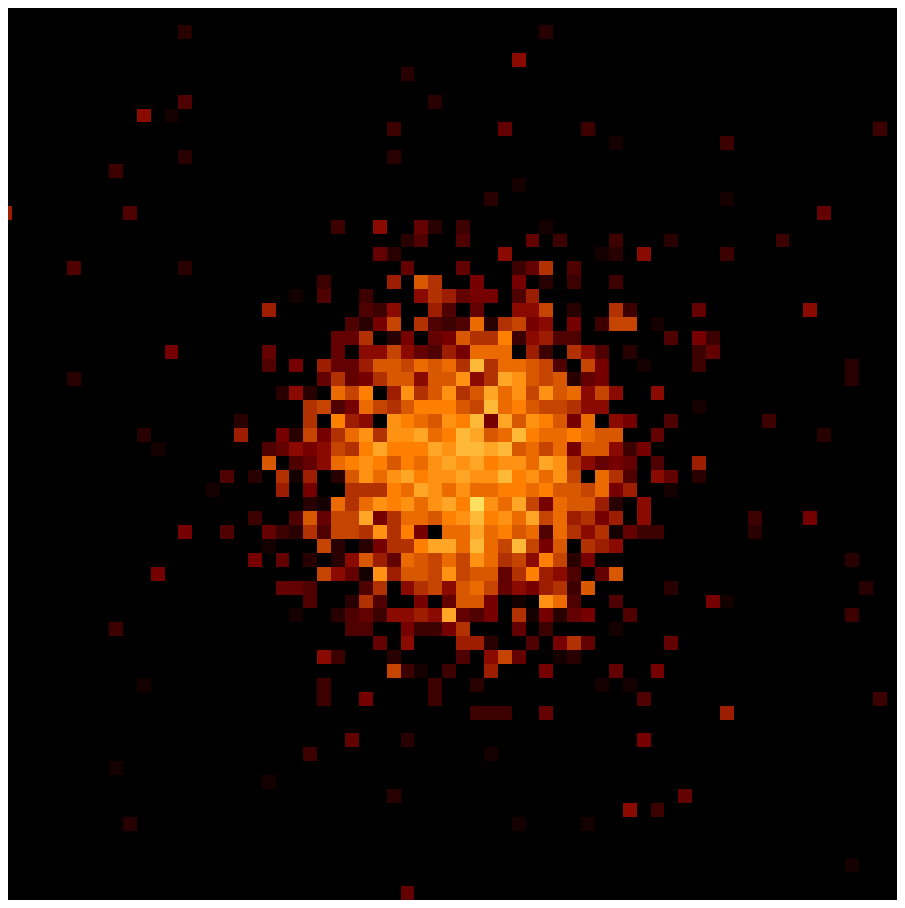}

\vspace{2mm}

\plotone{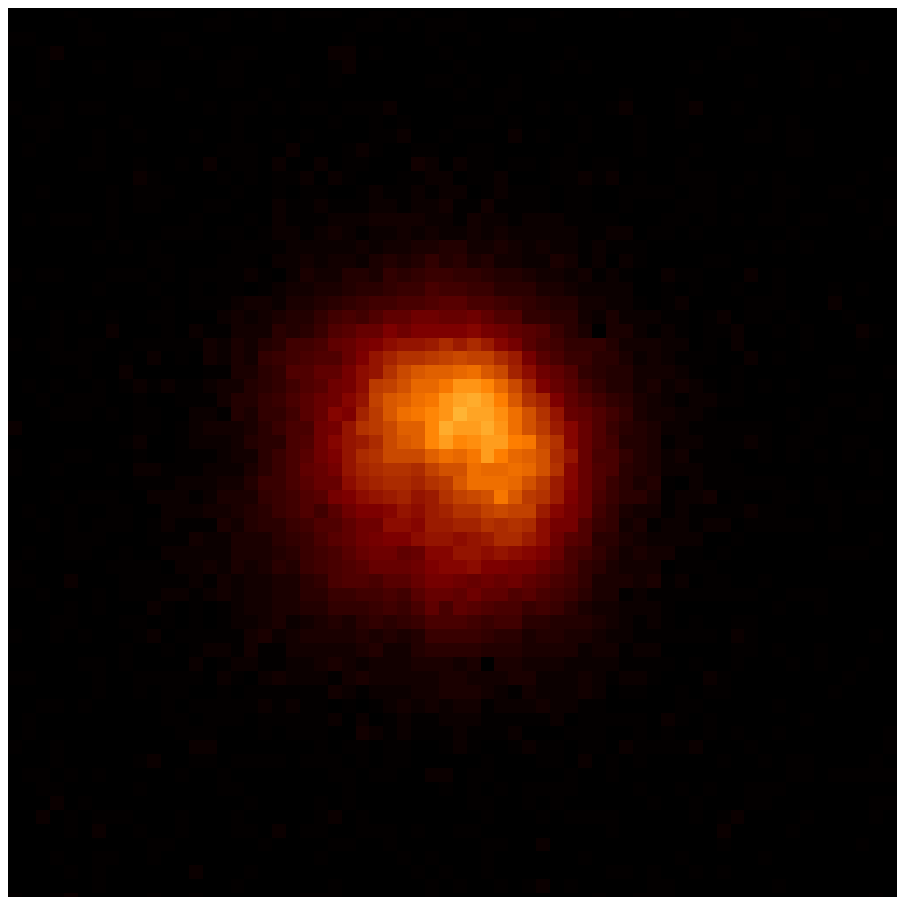}
\hspace{8mm}
\plotone{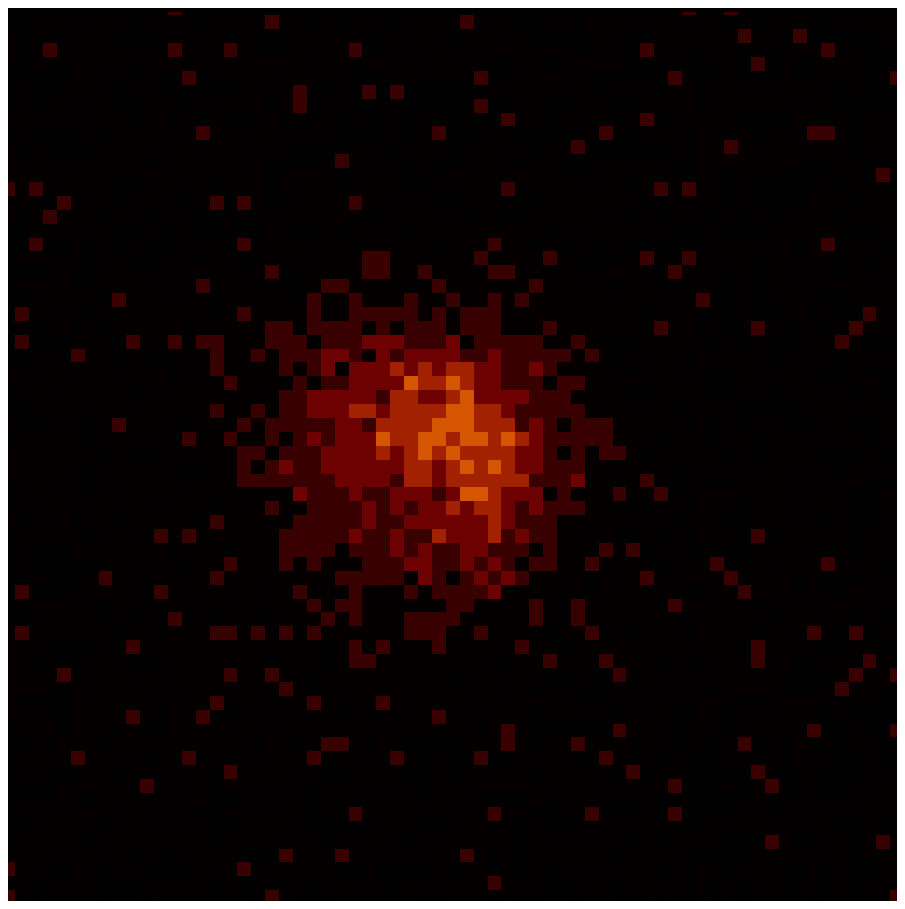}
\hspace{1mm}
\plotone{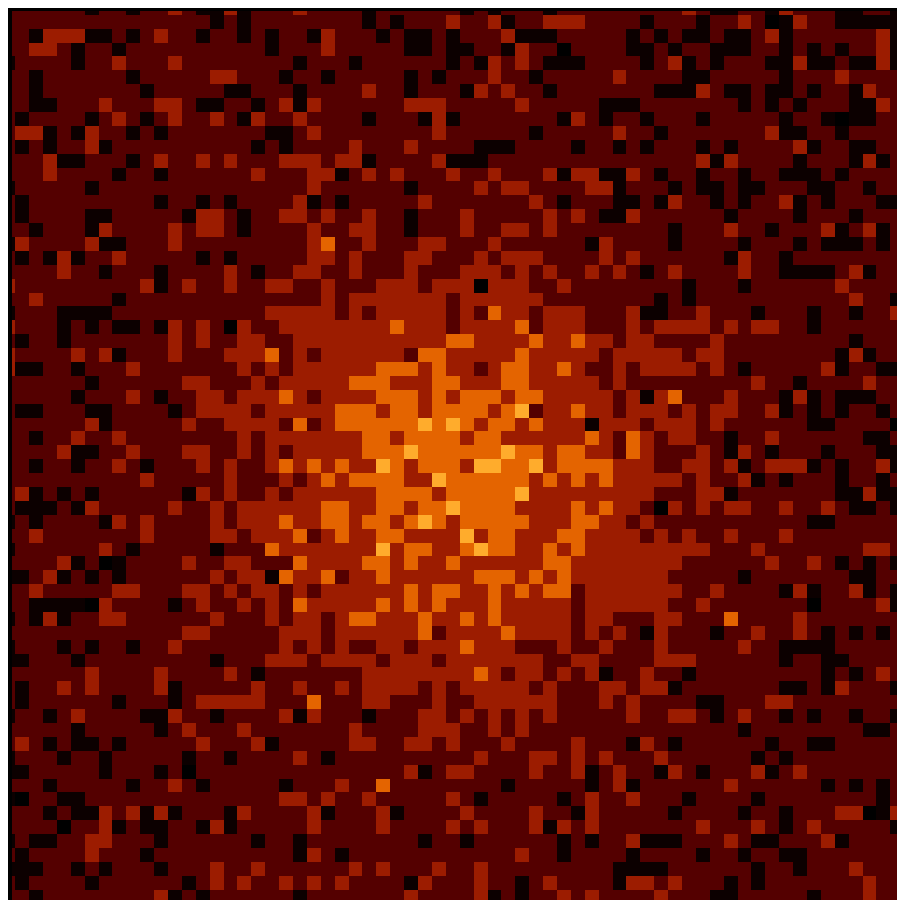}
\caption{Image sections 5.4\arcsec\ across corresponding to the
profiles drawn below:
of the tip-tilt star (left), and in UGC\,1347
the compact \hiiregion\ (centre) and nucleus (right);
for both open loop (lower) and closed loop on the laser guide star (upper).
Pixels are 0.08\arcsec\ across.
The LGS-corrected PSF is not perfect but does show significant
improvment beyond that achieved with tip-tilt alone.
\label{fig:ugc1347.psfs}}

\plotfiddle{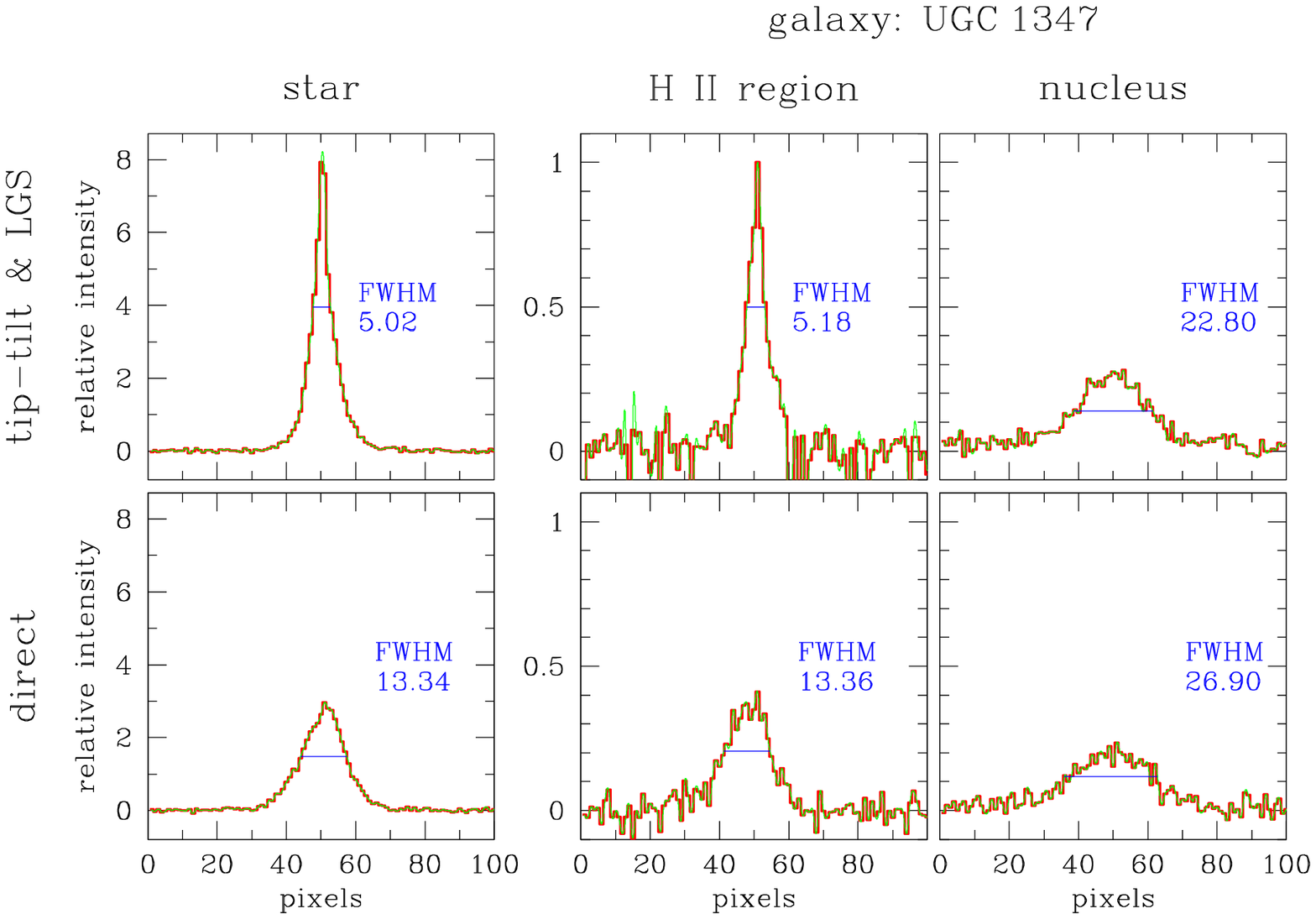}{9.5cm}{0}{70}{70}{-215}{-120}
\caption{Profiles of the star (left), and the compact \hiiregion\ and
nucleus of UGC\,1347 (right) in both open loop (lower) and closed loop
(upper).
FWHM are given in pixels, with a scale of 0.08\arcsec.
The peak intensity of the star increases by a factor of 2.5, as does
the unresolved \hiiregion, while there is almost no enhancement in the
nucleus showing it is completely resolved.
The laser was used to correct the high orders in these images, while
tip-tilt was determined from the star.
Tip-tilt alone produced little improvementin image quality.
\label{fig:ugc1347}}
\end{figure}

For UGC\,1347 we corrected tip-tilt using a V=11.8\,mag star 41$''$
away, and pointed the laser guide star midway between this and the
galaxy.
High order corrections were achieved using the LGS, also with 6
subapertures but at a lower frame rate of 50\,Hz (rejection
bandwidth of only 4\,Hz).
Figure~\ref{fig:ugc1347} shows for the star an increase in peak
intensity of 2.5, and improvement in resolution from 1.1$''$ to
0.4$''$.
Exactly similar enhancements are seen in the compact \hiiregion\ of the
galaxy 11\arcsec\ from the nucleus, but almost no change is seen in
the nucleus itself.
All the evidence for recent star formation in this galaxy can be
accounted for by the \hiiregion\ alone.

\subsection*{Galaxies in Abell Clusters}

As summarised in Table~\ref{tab:ugc}, both these galaxies have
projected positions close to the centre of the Abell\,262 cluster, and
both have large velocities with respect to the cluster average.
However, only UGC\,1344 is H{\sc i} deficient (Giovanelli \& Haynes
1985)\nocite{gio85}.

\begin{table}[ht]
\centering
\caption{Parameters for UGC\,1344 \& UGC\,1347 in Abell\,262\label{tab:ugc}}
\smallskip
\begin{tabular}{llllllll}
\hline

Galaxy & r/r$_{\rm Abell}$$^a$ & cz/\kms$^b$ & H{\sc i} Def. & 
M$_{\rm total}$/\msun & L$_{\rm FIR}$/\lsun & 
L$_{\rm K}$/\lsun (total)  & L$_{\rm K}$/\lsun (nuc.) \smallskip\\
\hline \hline
UGC\,1344  & 0.20 & 4155 & $>$0.78 & (0.8--1.2)$\times10^{10}$ &
$<4\times10^9$ & $3.3\times10^9$ & $5.6\times10^8$ \\
UGC\,1347  & 0.26 & 5543 & $-$0.07 & (0.5--1)$\times10^{11}$   & 
$1.2\times10^{10}$ & $1.9\times10^9$ & $2.6\times10^8$ \\

\hline
\end{tabular}

\smallskip
Note: $^a$ r$_{\rm Abell} = 1.44$\deg; $^b$ cz$_{\rm Abell} = 4830$\kms
\end{table}

H{\sc i} deficiency strongly correlates with distance from the cluster
centre, the central region being the zone of depletion;
and the fraction of deficient galaxies in a cluster correlates with the
observed 0.5--3.0\,keV X-ray luminosity of that cluster. 
The interpretation is that gas is stripped away due to environmental
effects of the inter-galactic medium as the galaxy crosses
the crowded central region, and the gas that is lost sinks to the
centre of the cluster gravitational potential, heating up and giving
rise to the hot intracluster gas observed in X-rays.
The mechanism by which the gas is stripped is uncertain, although both ram
pressure and evaporative stripping probably play some role.
Our results for the whole sample indicate a correlation between nucleus
core size with both Abell radius and H{\sc i} deficiency.
A narrow core size is indicative of a localised burst of star-formation,
occuring over a region perhaps 50-100\,pc compared with the typical
bulge size of 300\,pc.
The correlation suggests that a further effect is
induced as a galaxy crosses the centre of the cluster, namely star
formation.
This should not be surprising since near the cluster centre tidal
effects will be both much stronger (tidal force $F \propto 1/D^3$, D is
galaxy separation) and more common due to the higher galaxy density.
In this interpretation, the H{\sc i} deficient galaxy UGC\,1344 has
just passed the centre so that it has lost its gas and undergone a
burst of star formation;
UGC\,1347 is approaching the centre and has not yet been depleted of
its H{\sc i} nor has nuclear star formation been triggered.

\section{Conclusion}
\label{sec:conc}

We have presented an overview of recent results obtained with ALFA on
the 3.5-m telescope at Calar Alto during the summer semester of 1998
using the MPE 3D integral field spectrometer and $\Omega$-Cass,
including:
\begin{itemize}
\item
excellent performance on bright stars yielding, in the
K-band, diffraction limited resolution (0.14\arcsec) and high Strehl
ratios (40--60\%).
\item
the first spectroscopy at diffraction limited scales, clearly
distinguishing the spectra of both components of the
binary system HEI\,7 at a projected separation of 0.26\arcsec.
\item
spectroscopy of the Herbig Ae/Be star BD\,+40\deg\,4124 and JHK images
of an $80''\times80''$ field around it, resolving a binary system and a
circumstellar dust envelope.
\item
a laser guide star corrected image of the galaxy UGC\,1347 in the
Abell\,262 cluster which clearly highlights the difference between an
unresolved compact \hiiregion\ in the galaxy and its resolved nucleus.
\end{itemize}
ALFA is now beginning to show its true capabilites, and for some
observing programmes is able to compete effectively with other adaptive
optics sytems.
Observations with the laser guide star are still difficult, but
improvements which will be implemented during the autumn of 1998 should
stabilise closed loop operation and increase the observing efficiency.


\acknowledgments     
  
The MPIA/MPE team are grateful the Calar Alto staff for their help and
hospitality;
they also thank the rest of the ALFA team, as well as the 3D and
$\Omega$-Cass teams for their invaluable assistance.
RID acknowledges the support of the TMR (Training and Mobility of
Researchers) programme as part of the European Network for Laser Guide
Stars on 8-m Class Telescopes.

 
  \bibliography{/afs/mpa/home/davies/reference}   
  \bibliographystyle{/afs/mpa/data/irsi/v3/rid/styfiles/spie/spiebib}
  
  \end{document}